\documentclass[conference]{IEEEtran}
\IEEEoverridecommandlockouts

\usepackage{cite}
\usepackage{amsmath,amssymb,amsfonts}
\usepackage{algorithmic}
\usepackage{graphicx}
\usepackage{textcomp}
\usepackage{xcolor}
\usepackage{booktabs}
\usepackage{array}
\usepackage{soul}
\usepackage[caption=false,font=footnotesize]{subfig}  
\usepackage{graphicx}

\def\BibTeX{{\rm B\kern-.05em{\sc i\kern-.025em b}\kern-.08em
    T\kern-.1667em\lower.7ex\hbox{E}\kern-.125emX}}
\begin{document}

\title{Channel-Agnostic Semantic Compression for Bandwidth-Limited Visual Communication}

\author{
    \IEEEauthorblockN{
        Xuanhao Luo$^{1,*}$,
        Ruichen Gao$^{2,*}$,
        Zhizhen Li$^{1}$,
        Mingzhe Chen$^{3}$,
        Yuchen Liu$^{1}$
    }
    \IEEEauthorblockA{
        $^{1}$North Carolina State University, USA,
        $^{2}$New York University, USA,
        $^{3}$University of Miami, USA
    }
    \thanks{$^{*}$These authors contributed equally to this work.}
}

\maketitle

\begin{abstract}
Bandwidth-limited visual communication systems require efficient transmission of high-dimensional data under dynamic wireless conditions. Existing approaches either rely on joint source–channel coding, which tightly couples representation learning with channel models and lacks flexibility across varying environments, or adopt generative reconstruction techniques that may introduce semantically inconsistent outputs. In this paper, we propose RQ-NAC, a channel-agnostic semantic compression framework for visual communication. The proposed method leverages residual quantization to produce scalable discrete semantic representations, enabling fine-grained and predictable control over the rate-distortion tradeoff. To further enhance compression efficiency, we integrate an $n$-gram–driven arithmetic coding module that exploits contextual dependencies among latent indices for lossless entropy coding. Extensive experiments demonstrate that RQ-NAC\textsuperscript{1} achieves over 600× compression relative to uncompressed visual data while preserving high perceptual quality. The results show that our approach enables efficient, flexible, and reliable semantic transmission under bandwidth-constrained conditions.
\footnotetext[1]{{The source code of our proposed RQ-NAC pipeline is available at \textit{https://github.com/RickGao/RQ-NAC}.}}
\end{abstract}

\section{Introduction}
\label{sec:intro}

Recent bandwidth–limited visual transmission scenarios, such as vehicular perception \cite{chen2019cooper}, satellite imaging \cite{li2025adaorb}, and multi-modal wireless sensing \cite{li2026mmsense}, require the efficient delivery of high-dimensional visual data under highly constrained wireless links. These systems must frequently transmit camera or LiDAR images in real time, while channel conditions fluctuate due to mobility, interference, or environmental dynamics. Vehicular networks represent a particularly demanding instance of this challenge, where cooperative perception \cite{bista2018semantic}, collision warning \cite{luo2023clothoid}, and accident reporting \cite{zhong2023secure} rely on rapid dissemination of visual content under strict latency constraints \cite{gao2022communication}.
To make such high-volume visual data transmittable over bandwidth-limited wireless links, systems rely on codecs, which are compression and decompression algorithms that reduce data size by removing statistical redundancy.
However, conventional codecs such as JPEG and HEVC operate only at the pixel level and remain unaware of task semantics, leading them to allocate bits uniformly across the scene, including to task-irrelevant regions like background textures. In contrast, a semantic compressor focuses on preserving only the information that is essential for downstream understanding or decision-making~\cite{guo2024survey, yang2022semantic}, achieving higher compactness by abstracting redundant details. These characteristics make semantic-level compression appealing for a wide range of bandwidth-limited visual communication scenarios.

\begin{figure*}[t]
	\centerline{\includegraphics[scale=1.09]{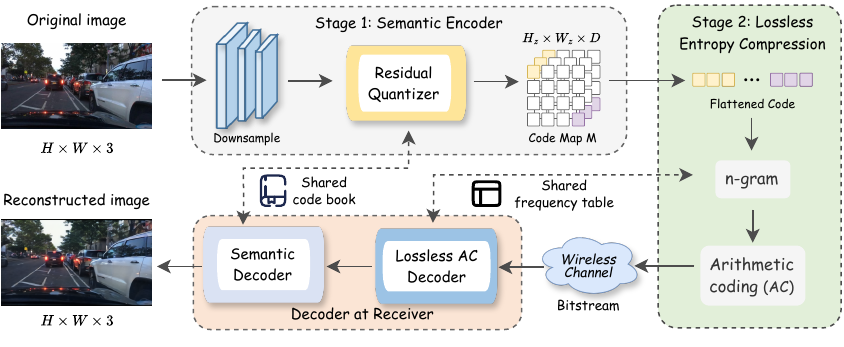}}
    \vspace{-3mm}
    \caption{Overview of the proposed RQ-NAC pipeline, consisting of a residual-quantized semantic encoder and a context-adaptive entropy coding module.} 
    \label{framework}
    \vspace{-3mm}
\end{figure*}

In general, a semantic communication system can be grouped into two major paradigms based on their transmission objectives: \textit{task-oriented} and \textit{reconstruction-oriented} \cite{qin2021semantic}.
Task-oriented methods transmit features optimized for specific tasks such as classification or detection, while reconstruction-oriented methods aim to recover semantically faithful signals at the receiver. Many recent works adopt deep joint source–channel coding (JSCC)~\cite{xu2023deep}, which jointly optimizes source representation and channel transmission under specific physical-layer settings.
However, such designs inherently \emph{couple} source and channel coding, making them highly dependent on channel assumptions (e.g., additive white Gaussian noise) and requiring costly retraining when channel conditions vary \cite{tang2026cache}. On the other hand, as channel quality degrades, JSCC systems may exhibit uncontrolled performance degradation \cite{bourtsoulatze2019deep}, which is undesirable for safety-critical applications where reliability and predictability are essential.
For the second paradigm, generative reconstruction approaches based on GANs and diffusion models reconstruct data by sampling from learned data priors, which often produce perceptually plausible but semantically inconsistent outputs, i.e., hallucinations. Such behavior undermines the reliability of reconstructed content, especially in scenarios where accurate semantic fidelity is crucial.

Motivated by these limitations, we take a different perspective and focus on \textbf{\textit{channel-agnostic}} semantic \textbf{\textit{source coding}}, aiming to learn compact and structured representations that enable predictable and controllable rate-distortion behavior through adaptive adjustment of the residual quantization depth, while being flexibly integrated with diverse channel conditions. We propose \textbf{RQ-NAC}, a two-stage semantic compression framework that integrates \underline{\textbf{R}}esidual \underline{\textbf{Q}}uantization with lossless \underline{\textbf{N}}-gram–driven \underline{\textbf{A}}rithmetic \underline{\textbf{C}}oding. In the first stage, a hierarchical residual quantization encoder generates discrete semantic representations with scalable capacity. In the second stage, an n-gram–based arithmetic coding module exploits contextual dependencies among latent indices to further reduce redundancy.
This two-stage pipeline effectively integrates learned semantic quantization with statistical entropy modeling, achieving both high-fidelity reconstruction and substantial bitrate reduction.
The main contributions of this work are summarized as follows:
\begin{itemize}
    \item We propose \textbf{RQ-NAC}, a novel channel-agnostic semantic compression framework that enables system-compatible visual communication by decoupling semantic representation from channel modeling.
    \item We develop a residual-quantization-based hierarchical semantic encoder that produces scalable discrete representations, where the bitrate is then controlled by the residual quantization depth, enabling predictable rate-distortion tradeoffs.
    \item We introduce an $n$-gram–driven arithmetic coding scheme that exploits contextual dependencies among discrete latent indices, bridging learned semantic representations with efficient entropy coding for lossless compression.
    \item Extensive experiments demonstrate that RQ-NAC achieves over $600\times$ compression while preserving high perceptual fidelity, and exhibits robust and stable reconstruction behavior under varying compression levels.
\end{itemize}

\section{Two-stage Adaptive Semantic Compression and Communication Pipeline}

As illustrated in Fig.~\ref{framework}, the proposed RQ-NAC framework consists of a residual-quantization–based semantic encoder followed by a lossless context-adaptive entropy coding module, forming a two-stage pipeline for scalable semantic compression and efficient transmission.

\subsection{Residual Quantization for Semantic Compression}
Unlike traditional vector quantization that performs a single-step codeword mapping, residual quantization hierarchically refines latent representations across multiple stages.
In contrast to variational autoencoders, which reconstruct data by sampling from continuous latent distributions, vector quantized variational autoencoder (VQ-VAE) \cite{van2017neural} and residual quantized variational autoencoder (RQ-VAE) \cite{lee2022autoregressive} adopt deterministic encoders with discrete latent representations. This deterministic formulation avoids the stochastic sampling process in generative models, thereby reducing the risk of semantically inconsistent reconstructions, while ensuring more reliable and faithful compression.
Building on this property, we repurpose RQ-VAE in the first stage as a standalone hierarchical semantic compression module, where the residual quantization produces scalable discrete representations suitable for communication.

Formally, let $E(\cdot)$ and $G(\cdot)$ denote the encoder and decoder, respectively. 
Given an input image $X \in \mathbb{R}^{H \times W \times 3}$, the encoder produces a latent feature map:
\begin{equation}
Z = E(X) \in \mathbb{R}^{H_z \times W_z \times n_z}, 
\end{equation}
where $H_z$ and $W_z$ are the spatial dimensions of the latent space, and $n_z$ is the feature dimensionality per spatial location. 
Each spatial feature vector $Z_{h,w} \in \mathbb{R}^{n_z}$ is quantized using residual quantization with depth $D$ and a shared codebook 
\begin{equation}
\mathcal{C} = \{ e(1), e(2), \ldots, e(K) \}, 
\end{equation}
where $e(k) \in \mathbb{R}^{n_z}$ denotes the $k$-th codeword, and $K$ is the codebook size. 

At stage $d$, the quantizer selects a codeword $e(k^{(d)}_{h,w})$ from the
codebook and refines the quantized approximation:
\begin{equation}
    \hat{Z}^{(d)}_{h,w} = \sum_{i=1}^{d} e(k^{(i)}_{h,w}), 
    \quad
    \hat{Z}_{h,w} = \hat{Z}^{(D)}_{h,w},
\end{equation}
where $d \in \{1,\ldots,D\}$ denotes the residual quantization depth.
The selected codeword indices are
\[
    M_{h,w} = \big(k^{(1)}_{h,w}, \ldots, k^{(D)}_{h,w}\big),
\]
which constitutes the discrete symbols generated during quantization. Collectively, 
these index sequences form the code map 
$M \in [K]^{H_z \times W_z \times D}$, which serves as the compact discrete 
representation transmitted over the wireless channel.
At the receiver side, the transmitted code map $M$ is used to reconstruct the 
quantized latent map $\hat{Z} \in \mathbb{R}^{H_z \times W_z \times n_z}$, 
which is subsequently decoded to obtain the final image $ \hat{X} = G(\hat{Z})$.

The overall training objective combines a reconstruction term and a multi-level commitment term as:
\begin{equation}
    \mathcal{L} 
    = \|X - \hat{X}\|_2^{2}
    + \beta \sum_{d=1}^{D} \|Z - \text{sg}[\hat{Z}^{(d)}]\|_2^{2},
\end{equation}
where $\text{sg}[\cdot]$ denotes the stop-gradient operator, and $\beta$ is a balancing coefficient. 
Following the RQ-VAE implementation, we update the codebook using exponential moving average (EMA) of the clustered encoder outputs instead of gradient-based updates. This approach stabilizes training and prevents codebook collapse. Additionally, perceptual loss and patch-level adversarial loss are incorporated to enhance visual fidelity.  This hierarchical quantization structure refines the latent representation level by level while sharing the same compact codebook across all depths, producing a discrete index tensor $M$.
We then flatten the index tensor $M$ into a one-dimensional sequence $\mathbf{s} = (s_1, s_2, \dots, s_N)$ with length $N = H_z W_z D$. This sequence serves as the input to the entropy coding stage as follows.

\subsection{Context-adaptive Arithmetic Coding}
After the semantic compression stage, the residual quantization encoder outputs a sequence of discrete indices 
$\mathbf{s}$. 
Although already compact, this sequence still contains statistical redundancy due to the strong correlation among neighboring indices. In the second stage, we further remove statistical redundancy by applying a context-adaptive lossless entropy coding scheme.

From an information-theoretic view, the minimum achievable code length of this sequence is its conditional entropy:
\begin{equation}
    H(S \mid C) = -\sum_{s_t,\mathbf{c}_t} p(s_t,\mathbf{c}_t)\,\log p(s_t \mid \mathbf{c}_t),
\end{equation}
where $p(s_t,\mathbf{c}_t)$ is the joint distribution of the symbol and its context.
In practice, arithmetic coding uses an estimated model $q(\cdot)$, leading to an expected coding rate given by the cross-entropy $ H(p,q) = \mathbb{E}_{p(s_t,\mathbf{c}_t)}\!\big[-\log q(s_t \mid \mathbf{c}_t)\big].$
Thus, improving the accuracy of $q$ directly tightens the gap between the achievable rate and the entropy bound.

To better approximate $p(\cdot)$ and capture local dependencies, we employ an $n$-gram–based probabilistic model.  
Although predictive autoregressive models such as LLMs or transformer-based entropy models are used in natural language \cite{valmeekam2023llmzip} and byte-level \cite{luo2025rank, luo2025unified} compression, their sequential inference is too slow for real-time visual data transmission.
Thus, the $n$-gram model introduces low computational overhead while still capturing essential dependencies for entropy coding.
Given a context $\mathbf{c}_t = (s_{t-n+1}, \dots, s_{t-1})$, the conditional probability of the next symbol is estimated by maximum likelihood:
\begin{equation}
    \hat{P}(s_t \mid \mathbf{c}_t) = 
    \frac{N(\mathbf{c}_t, s_t)}{N(\mathbf{c}_t)},
\end{equation}
where $N(\mathbf{c}_t, s_t)$ denotes the number of occurrences of the $(\mathbf{c}_t, s_t)$ pattern in the training corpus, and $N(\mathbf{c}_t)$ is the number of occurrences of the context $\mathbf{c}_t$.  
By conditioning on $\mathbf{c}_t$, the $n$-gram estimator provides a more faithful approximation of the true sequence statistics, thereby reducing the cross-entropy term $H(p,q)$ and improving lossless compression efficiency.

However, when certain $n$-gram combinations are unseen during training, direct estimation may yield zero probabilities, which can collapse the arithmetic coding (AC) interval. 
To mitigate this issue, we apply additive smoothing \cite{chen1999empirical} that assigns a small nonzero probability to unseen symbols. 
The smoothed conditional probability is defined as:
\begin{equation}
    \tilde{P}(s_t \mid \mathbf{c}_t) = 
    \frac{N(\mathbf{c}_t, s_t) + \alpha}{N(\mathbf{c}_t) + \alpha K},
\end{equation}
where $\alpha$ is a small positive smoothing coefficient ($0 < \alpha \leq 1$) and $K$ is the codebook size.  
This approach ensures that every possible symbol maintains a nonzero probability, thereby preventing numerical instability and guaranteeing robust encoding and decoding across all contexts.
In our framework, the $n$-gram estimator $\tilde{P}(s_t \mid \mathbf{c}_t)$ serves as the conditional probability model $q(s_t \mid \mathbf{c}_t)$ used by AC.
The $n$-gram frequency table is constructed offline and shared between encoder and decoder, so no additional side information needs to be transmitted.

After obtaining the probability distribution from the n-gram model, AC progressively narrows an interval $[l,h)$ to encode the entire sequence.
At step $t$, the interval is updated according to the cumulative distribution function $F(\cdot \mid \mathbf{c}_t)$ of $\hat{P}(\cdot \mid \mathbf{c}_t)$:
\begin{equation}
\begin{aligned}
    l_t &= l_{t-1} + (h_{t-1} - l_{t-1}) \cdot F(s_t \mid \mathbf{c}_t), \\
    h_t &= l_{t-1} + (h_{t-1} - l_{t-1}) \cdot F(s_t+1 \mid \mathbf{c}_t).
\end{aligned}
\end{equation}
where:
\begin{equation}
    F(k \mid \mathbf{c}_t) = \sum_{j < k} \tilde{P}(j \mid \mathbf{c}_t).
\end{equation}
Through this iterative refinement, the interval $[l, h)$ becomes increasingly \textit{narrow} as more symbols are encoded. After processing all $N$ indices, the final interval $[l_N, h_N)$ uniquely represents the entire sequence, and any real number within this range can serve as the compact binary representation transmitted to the receiver.

At the receiver side, the decoder first loads the shared $n$-gram frequency table, which contains the conditional probability model used during encoding. Using this model, the arithmetic decoder reconstructs the latent index sequence in a sequential manner: starting from the initial interval $[0,1)$, it repeatedly finds the sub-interval whose probability range contains the encoded value and updates the interval accordingly. This process continues until all indices are recovered, producing the exact same discrete code map as the encoder. The recovered indices are then mapped to codewords using the shared codebook to obtain the quantized latent representation $\hat{Z}$, which is finally fed into the semantic decoder to generate the reconstructed data.

\begin{figure*}[t]
	\centerline{\includegraphics[scale=0.15]{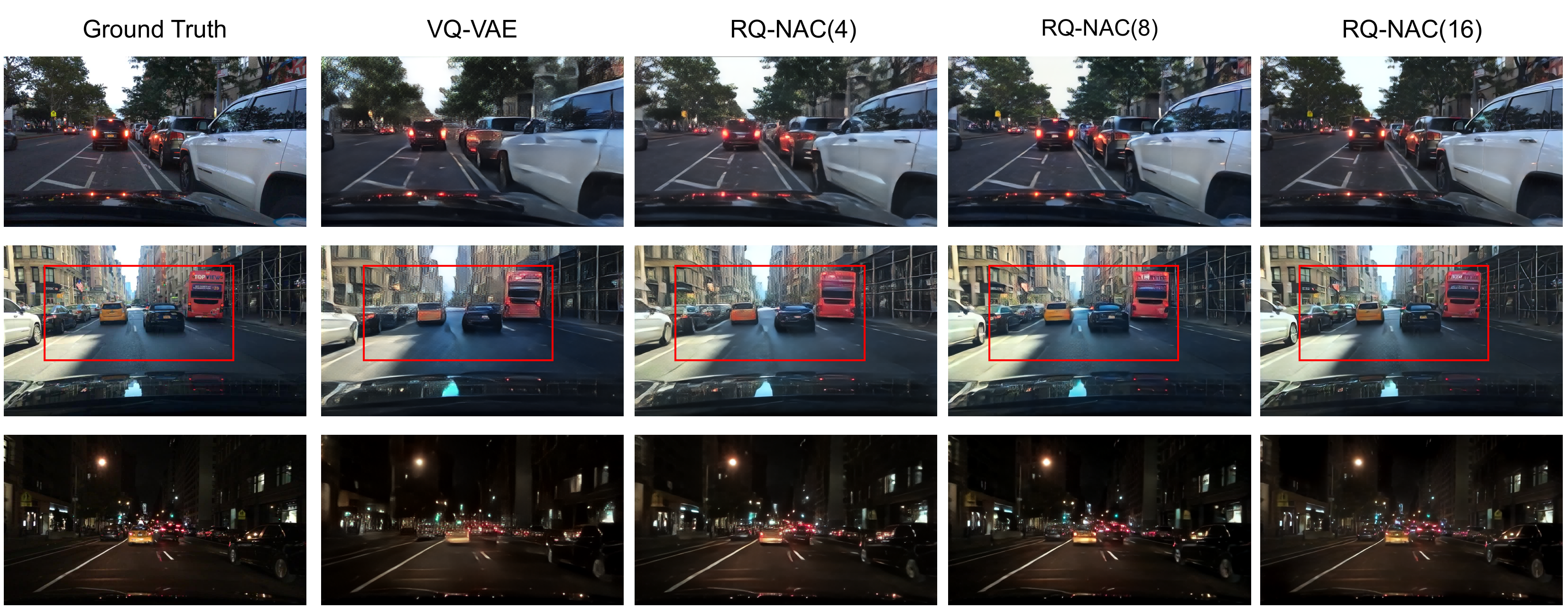}}
    \caption{Qualitative comparison of reconstructed images produced by different compression models. 
    The proposed RQ-NAC progressively improves visual fidelity as the residual quantization depth increases, recovering sharper edges, clearer object contours, and more stable scene structures.
    } 
    \label{iov}
    \vspace{-4mm}
\end{figure*}

\subsection{Model Adaptation for Efficient Communication}
\label{subsec:adaptive_comm}

To enable adaptive communication under varying channel bandwidth and latency constraints, the proposed RQ-NAC framework supports dynamic adjustment of two key components: the residual quantization \textbf{depth} and the $n$-gram \textbf{order} in the entropy model. 
The hierarchical residual quantization structure provides multiple encoding depths, where each additional level progressively refines the residual error of the previous one. As a result, deeper quantization captures finer semantic details and achieves higher perceptual reconstruction quality, but incurs larger latent indices and increased bitrate. Conversely, shallower models achieve higher compression efficiency with shorter bitstreams, at the cost of lower reconstruction fidelity due to the limited representational capacity. 

Similarly, the $n$-gram–based entropy model introduces a trade-off between contextual modeling accuracy, computational complexity, and statistical reliability. In theory, increasing $n$ should improve compression efficiency, as longer contexts reduce the conditional entropy by capturing higher-order dependencies among latent indices.
However, in practice, the available data are limited while the context space grows exponentially with $n$, causing severe data sparsity and unreliable probability estimation.
Many rare or unseen contexts must be smoothed toward nearly uniform distributions, which increases empirical entropy and reduces actual compression efficiency.
Moreover, the computational and memory cost of constructing and querying an $n$-gram model also scales exponentially with $n$, making high-order models impractical for real-time encoding and decoding.
Therefore, a moderate $n$ provides the best trade-off between contextual expressiveness, computational cost, and compression performance.

\section{Experiments and Evaluation Results}

\subsection{Experimental Setup and Baselines}
\label{setup}

We evaluate RQ-NAC in a bandwidth-limited vehicular communication setting, where onboard cameras continuously capture road scenes that must be efficiently transmitted over resource-constrained wireless links. We adopt 20,000 (20k) images from the 100K Vehicle Dashcam Image Dataset~\cite{dashcam_dataset} for training and 1,000 images for testing. To reduce computational cost, we randomly sample a subset of 5,000 (5k) images from the training set for most experiments, while keeping the test set fixed. To evaluate model generalization in out-of-distribution scenarios, we also conduct a case study on 1,000 frames extracted from the Car Crash Dataset~\cite{car_crash_dataset}, which contains real-world accident scenes \textit{not} seen during training.
We compare our proposed RQ-NAC models with different quantization depths, the standard VQ-VAE, and the traditional non-learned codecs HEVC and JPEG. The codebook size is set to 2,048 for all models. Input images, originally sized $1,280{\times}720$, are zero-padded along the height to $1,280{\times}736$ to ensure the dimension is divisible by 32, which facilitates downsampling operations in the encoder.
All models are trained on eight A100 GPUs using the NVIDIA Saturn Cloud platform.

\begin{table*}[t]
\centering
\caption{Performance comparison under different compression configurations. 
``$\pm$'' indicates the standard deviation computed over the test set.
\textbf{Bold} and \underline{underlined} values indicate the best and second-best results among \textbf{learned} models, respectively. 
HEVC* and JPEG* are \textbf{non-learned} codecs included as reference baselines.}
\label{tab:compression_results}
\begin{tabular}{c|cccc|cc|c}
\toprule
\textbf{Configuration} & \textbf{SSIM$\uparrow$} & \textbf{PSNR (dB)$\uparrow$} & \textbf{LPIPS$\downarrow$} & \textbf{FID$\downarrow$} & \textbf{BPP$\downarrow$} & \textbf{CR$\uparrow$} (w/o AC) & \textbf{CR$\uparrow$} (w/ AC)\\

\midrule
VQ-VAE-5k      & 0.7047 $\pm$ 0.1028 & 22.42 $\pm$ 2.89 & 0.2667 $\pm$ 0.0758 & 39.1660 & \textbf{0.01098} & \textbf{2,185.61} & \textbf{2,166.38} \\
\midrule
RQ-NAC(4)-5k     & 0.7400 $\pm$ 0.1059 & 23.57 $\pm$ 2.62 & 0.2263 $\pm$ 0.0692 & 32.8649 & \underline{0.04392} & \underline{546.40} & \underline{671.09} \\
RQ-NAC(8)-5k     & 0.7637 $\pm$ 0.0874 & 24.80 $\pm$ 3.01 & 0.1821 $\pm$ 0.0579 & 26.3956 & 0.08784 & 273.20 & 356.25 \\
RQ-NAC(16)-5k    & \textbf{0.7832} $\pm$ 0.0654 & \textbf{25.00} $\pm$ 2.26 & \textbf{0.1519} $\pm$ 0.0424 & \underline{24.8398} & 0.17569 & 136.60 & 179.32 \\
\midrule
RQ-NAC(4)-20k & \underline{0.7779} $\pm$ 0.0858 & \underline{24.92} $\pm$ 3.29 & \underline{0.1541} $\pm$ 0.0451 & \textbf{22.4336} & 0.04392 & 546.40 & 627.81 \\
\midrule
JPEG* (original dataset)  & — & — &—  & —
&  0.5107 & 47.00 & —\\
HEVC* (reference)  & 0.9584 ± 0.0101 & 36.67 ± 2.29 & 0.0762 ± 0.0246 & 6.7903
 & 0.2105 &  114.03 & —\\
\bottomrule
\end{tabular}
\vspace{-4mm}
\end{table*}

\textbf{Reconstruction metrics.} To evaluate the perceptual quality of reconstructed images, we adopt four data reconstruction metrics in different dimensions. 
The Structural Similarity Index Measure (SSIM) measures the structural and luminance similarity between the reconstructed and ground-truth images. 
The Peak Signal-to-Noise Ratio (PSNR) quantifies pixel-level reconstruction accuracy based on mean squared error (MSE). 
The Learned Perceptual Image Patch Similarity (LPIPS) evaluates perceptual similarity using deep feature representations. 
Finally, the Fréchet Inception Distance (FID) measures the distributional distance between reconstructed and ground-truth images in the feature space.

\textbf{Compression metrics.} To evaluate the compression efficiency and data compactness, we employ bits per pixel (BPP) and compression ratio (CR). 
The BPP represents the average number of bits required to encode each pixel in the compressed image, i.e., 
$\text{BPP} = \frac{b_{\text{comp}}}{H \times W}$,
where \(b_{\text{comp}}\) denotes the total number of bits in the compressed bitstream, and \(H\) and \(W\) are the image height and width, respectively.  
The CR is defined as the ratio between the uncompressed and compressed data sizes, i.e., $\text{CR} = \frac{\text{S}_{\text{orig}}}{\text{S}_{\text{comp}}}$,  where $\text{S}_{\text{orig}}$ and $\text{S}_{\text{comp}}$ denote the sizes of the original and compressed data, respectively.
We also report the native compression ratio of the dataset JPEG files, but no reconstruction metrics are provided for JPEG since they are the original images rather than reconstructed outputs.

\subsection{Performance of Semantic Compression}

We compare the proposed RQ-NAC with traditional codecs and baselines, as shown in Table~\ref{tab:compression_results}. We include VQ-VAE as a learning-based baseline, which can be viewed as a single-stage quantization model.
Overall, we observe a clear trend that increasing the quantization depth consistently improves reconstruction quality across all perceptual and distortion metrics. 
For example, RQ-NAC(16) achieves the best performance among learned models, reaching an SSIM of 0.7832, PSNR of 25.00 dB, and LPIPS of 0.1519, significantly outperforming the VQ-VAE baseline (SSIM 0.7047, PSNR 22.42 dB). 
This improvement comes at the cost of reduced compression efficiency, where the compression ratio decreases from 546.40× at a depth of 4 to 136.60× at a depth of 16, reflecting the trade-off between reconstruction quality and bitrate.
Fig.~\ref{iov} qualitatively confirms these trends, where deeper models produce sharper details and more faithful reconstructions.

To further investigate the impact of training data size, we evaluate a model with quantization depth 4 trained on an expanded dataset of 20k samples. 
As shown in Table~\ref{tab:compression_results}, the RQ-NAC(4)-20k model achieves comparable reconstruction performance to the much deeper RQ-NAC(16)-5k variant (e.g., SSIM 0.7779 vs. 0.7832), indicating that increasing the amount of training data can enhance the model’s representational capacity and improve the reconstruction quality with lower bitrate overhead.  
Unless otherwise specified, all other models are trained on the 5k subset described in Section~\ref{setup}.
At the same time, HEVC (CRF = 28), as a hand-crafted algorithm, achieves the highest reconstruction fidelity. However, it operates at a much higher bitrate, resulting in a compression ratio of only 114.03. In contrast, our RQ-NAC(4) achieves a compression ratio of 546.40, representing a 379.1\% improvement while maintaining reasonable perceptual quality.

\subsection{Performance of Entropy-based Lossless Compression}

\begin{figure}[t]
  \centering
  \subfloat[Impact of $n$-gram order on compression ratio. \label{fig:ngram}]{
    \includegraphics[width=0.46\linewidth]{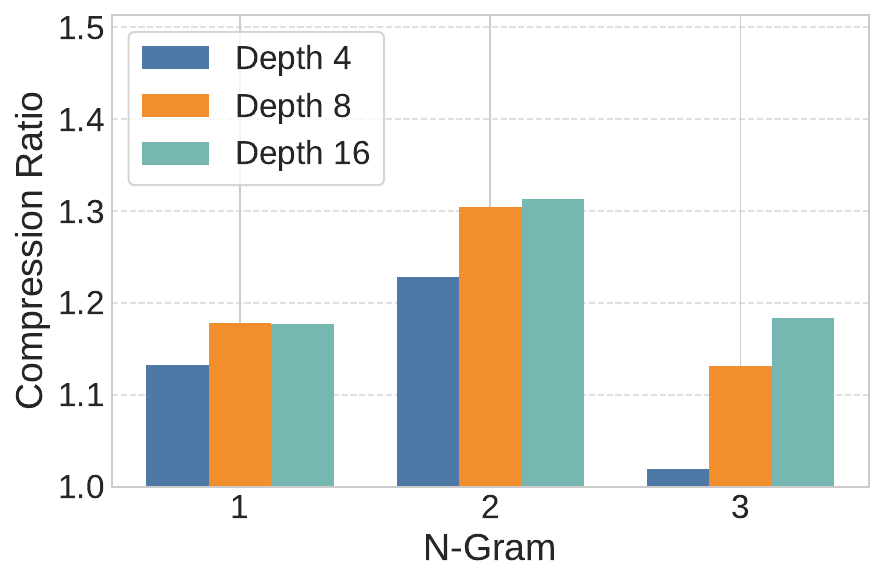}
  }
  \hfill
  \subfloat[Impact of coefficient $\alpha$ on compression ratio. \label{fig:alpha}]{
    \includegraphics[width=0.46\linewidth]{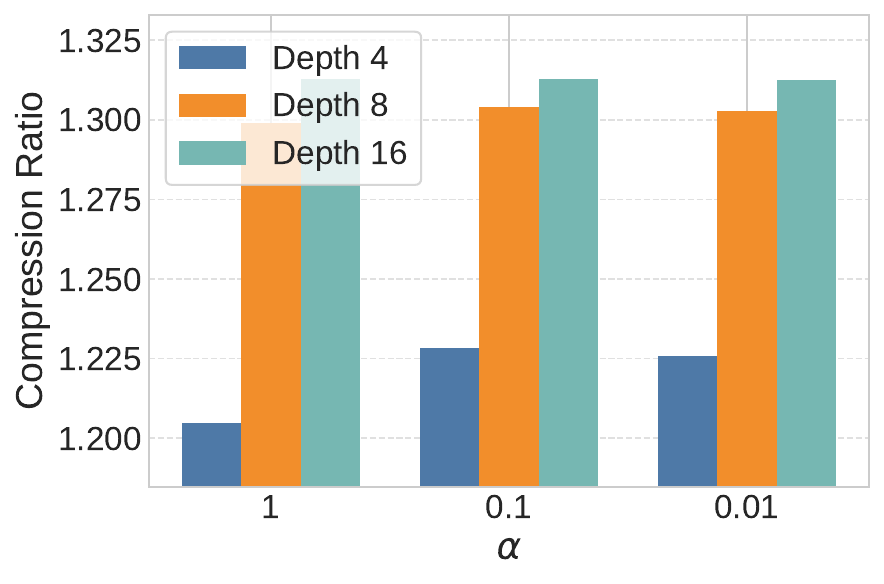}
  }
  \caption{Parametric study on compression ratio under different quantization depths.}
  \label{fig:compression-ratio}
  \vspace{-4mm}
\end{figure}

We evaluate the effectiveness of the proposed context-adaptive entropy coding model under different $n$-gram orders and quantization depths. 
Fig.~\ref{fig:ngram} presents the resulting stage-2 compression ratios across different quantization depths.
We observe that even without the contextual modeling ($n=1$), applying AC directly yields a notable compression ratio of approximately $1.15\times$, demonstrating the redundancy present in the latent space. 
When incorporating a 2-gram context model, the compression ratio improves significantly to around $1.3\times$, indicating that short-range dependencies between adjacent code symbols can be further leveraged to enhance compression performance.
However, the 3-gram case leads to a degradation in compression efficiency because the estimated empirical frequencies $\tilde{P}(s_t \mid \mathbf{c}_t)$ deviate from the true probabilities $P(s_t \mid \mathbf{c}_t)$ under \textit{limited data}. 
The exponentially growing context space causes such data sparsity, leading to unreliable and nearly uniform estimates that increase entropy and reduce compression efficiency.
In theory, higher-order $n$-grams can yield better compression as they more accurately model local dependencies, but require significantly larger datasets to avoid context sparsity and maintain reliable probability estimates. 
Besides, both the storage and lookup costs for frequency tables grow exponentially with $n$, which can quickly become impractical for resource-constrained applications.

To mitigate sparsity, we evaluate additive smoothing with $\alpha \in \{1, 0.1, 0.01\}$ under the 2-gram model. As shown in Fig.~\ref{fig:alpha}, varying $\alpha$ results in only marginal compression differences across quantization depths, with $\alpha = 0.1$ slightly outperforming others at depth of 4 and 8. We therefore adopt $\alpha = 0.1$ as the default, and all “w/ AC” results in Table~\ref{tab:compression_results} use this configuration. 
The VQ-VAE baseline exhibits degraded performance with n-gram modeling because its relatively short latent sequences limit the effectiveness of context-based estimation and may introduce estimation bias. In contrast, 2-gram modeling is most effective for our RQ-NAC, achieving up to \textbf{671.09×} compression at depth of 4 with high perceptual quality.

\begin{figure}[t]
	\centerline{\includegraphics[scale=0.7]{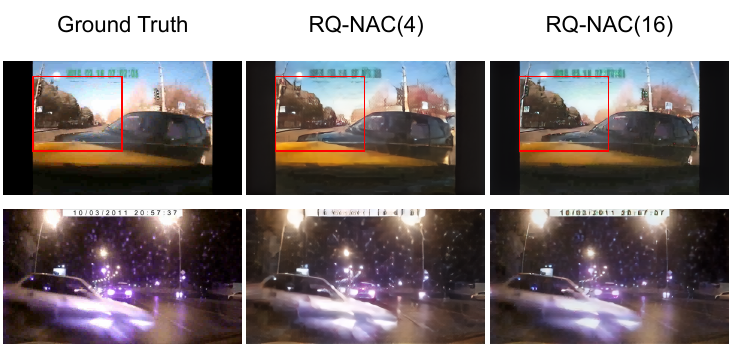}}
    \caption{Case study on unseen accident scenes. RQ-NAC maintains strong reconstruction quality under distribution shift, with deeper quantization providing clearer textures and sharper scene details.} 
    \label{case}
\end{figure}

\subsection{Case Study: Car Crash Scenario}

To evaluate the generalization ability of our RQ-NAC, we conduct a case study using the Car Crash Dataset, which contains real-world accident scenes that were not seen during training. We directly apply two RQ-NAC models pretrained on our original vehicular dataset—one with depth 4 and the other with depth 16—without any fine-tuning.
As shown in Fig.~\ref{case}, our models maintain strong reconstruction performance. The depth-4 model is able to recover all essential scene elements, such as vehicle outlines, traffic lights, and road structures, which are sufficient for downstream perception tasks. In contrast, the depth-16 model reconstructs finer details like object textures and edge sharpness, providing a more faithful visual representation.
Table~\ref{tab:case} further quantifies the perceptual quality. While both models experience a slight performance drop compared to their results on the original test set, the degradation is modest, demonstrating high-fidelity reconstruction under distribution shift.
These results demonstrate the generalization capability of our RQ-NAC models. Even without exposure to this specific car-crash domain, the pretrained models preserve semantic integrity and maintain robust reconstruction quality.

\section{Conclusion}

This paper presents a two-stage adaptive semantic  compression framework for bandwidth-limited visual communication systems. The proposed RQ-NAC integrates residual quantization with an n-gram–based arithmetic coder to achieve high compression efficiency while maintaining semantic fidelity. 
Experiments show that our framework achieves more than $600\times$ compression relative to the uncompressed data size while preserving strong perceptual quality, and generalizes well to unseen accident scenarios.

\section*{Acknowledgment}
This research was supported by the National Science Foundation through Award CNS-2440756, CNS-2312138, and NVIDIA Academic Program Grant.

\begin{table}[t]
\centering
\caption{Quantitative results of the case study.}
\label{tab:case}
\begin{tabular}{c|cccc}
\toprule
\textbf{Configuration} & \textbf{SSIM$\uparrow$} & \textbf{PSNR (dB)$\uparrow$} & \textbf{LPIPS$\downarrow$} & \textbf{FID$\downarrow$}  \\

\midrule
RQ-NAC(4)      & 0.6764 & 21.73 & 0.2752 & 72.2823  \\
\midrule
RQ-NAC(16)     & 0.7369 & 23.07 & 0.2067 & 54.3657 \\
\bottomrule
\end{tabular}
\end{table}

\bibliographystyle{IEEEtran}
\bibliography{references}

\end{document}